\documentclass[12pt]{article}
        \newdimen\eqskip
        \newdimen\txtskip
        \baselineskip=\txtskip
        \newdimen\mysep                
        \newdimen\hmysep
\begin{document}
\renewcommand{\thefootnote}{\fnsymbol{footnote}}
  \newcommand{\ccaption}[2]{
    \begin{center}
    \parbox{0.85\textwidth}{
      \caption[#1]{\small{{#2}}}
      }
    \end{center}
    }
\newcommand{\BS}{\bigskip}
\def    \be             {\begin{equation}}
\def    \ee             {\end{equation}}
\def    \beq             {\begin{equation}}
\def    \eeq             {\end{equation}}
\def    \ba             {\begin{eqnarray}}
\def    \ea             {\end{eqnarray}}
\def    \beqn           {\begin{eqnarray}}
\def    \eeqn           {\end{eqnarray}}
\def    \beeq           {\begin{eqnarray}}
\def    \eeeq           {\end{eqnarray}}
\def    \nn             {\nonumber}
\def    \=              {\;=\;}
\def    \frac           #1#2{{#1 \over #2}}
\def    \rd             {\rm d}
\def    \ret            {\\[\eqskip]}
\def    \ie             {{\em i.e.\/} }
\def    \eg             {{\em e.g.\/} }
\def \lsim{\mathrel{\vcenter
     {\hbox{$<$}\nointerlineskip\hbox{$\sim$}}}}
\def \gsim{\mathrel{\vcenter
     {\hbox{$>$}\nointerlineskip\hbox{$\sim$}}}}
\def    \bentarrow      {\:\raisebox{1.1ex}{\rlap{$\vert$}}\!\rightarrow}
\def    \rd             {{\mathrm d}}    
\def    \Im             {{\mathrm{Im}}}  
\def    \bra#1          {\mbox{$\langle #1 |$}}
\def    \ket#1          {\mbox{$| #1 \rangle$}}
\def    \to             {\rightarrow} 

\def    \kev            {\mbox{$\mathrm{keV}$}}
\def    \mev            {\mbox{$\mathrm{MeV}$}}
\def    \gev            {\mbox{$\mathrm{GeV}$}}


\def    \mq             {\mbox{$m_Q$}}  
\def    \mt             {\mbox{$m_t$}}  
\def    \mb             {\mbox{$m_b$}}  
\def    \mqq            {\mbox{$m_{Q\bar Q}$}}
\def    \mqqsq          {\mbox{$m^2_{Q\bar Q}$}}
\def    \pt             {\mbox{$p_T$}}
\def    \ptsq           {\mbox{$p^2_T$}}

\newcommand     \MSB            {\ifmmode {\overline{\rm MS}} \else 
                                 $\overline{\rm MS}$  \fi}
\def    \muf            {\mbox{$\mu_{\rm F}$}}
\def    \mug            {\mbox{$\mu_\gamma$}}
\def    \mufsq          {\mbox{$\mu^2_{\rm F}$}}
\def    \mur            {{\mbox{$\mu_{\rm R}$}}}
\def    \mursq          {\mbox{$\mu^2_{\rm R}$}}
\def    \mul            {{\mu_\Lambda}}
\def    \mulsq          {\mbox{$\mu^2_\Lambda$}}

\def    \bzero          {\mbox{$b_0$}}
\def    \as             {\ifmmode \alpha_S \else $\alpha_S$ \fi}
\def    \asb            {\mbox{$\alpha_S^{(b)}$}}
\def    \assq           {\mbox{$\alpha_S^2$}}
\def \oacube {\mbox{$ O(\alpha_S^3)$}}
\def \oafour {\mbox{$ O(\alpha_S^4)$}}
\def \oatwo {\mbox{$ O(\alpha_S^2)$}}
\def \oas   {\mbox{$ O(\alpha_S)$}}
\def\asp{{\alpha_S}\over{\pi}}

\def\slash#1{{#1\!\!\!/}}
\def\rt1{\raisebox{-1ex}{\rlap{$\; \rho \to 1 \;\;$}}
\raisebox{.4ex}{$\;\; \;\;\simeq \;\;\;\;$}}
\def\ltap{\raisebox{-.5ex}{\rlap{$\,\sim\,$}} \raisebox{.5ex}{$\,<\,$}}
\def\gtap{\raisebox{-.5ex}{\rlap{$\,\sim\,$}} \raisebox{.5ex}{$\,>\,$}} 

\def\ylim{\raisebox{-1.5ex}{\rlap{$\, y \to 0 \,$}} 
\raisebox{.1ex}{$\;\;\; \longrightarrow \;\;$}}

\def\yeq{\raisebox{-1.5ex}{\rlap{$\, y \to 0 \,$}} 
\raisebox{.1ex}{$\;\;\;\; = \;\;\;\;$}} 

\def \ee  {e^+e^-}
\def\GE{\gamma_E}
\def\half{\frac{1}{2}}
\def\b0{\beta_0}
\def\naive{na\"{\i}ve}
\def\cm{{\cal M}}
\def\bom#1{\mbox{\bf{#1}}}
\begin{titlepage}
\nopagebreak       {\flushright{
        \begin{minipage}{5cm}
        CERN-TH/2003-125\\
	Bicocca-FT-03-17\\
        {\tt hep-ph/0307035}\\
        \end{minipage}        }

}
\vfill
\begin{center}
{\LARGE { \bf \sc 
SUDAKOV RESUMMATION \\
OF MULTIPARTON \\
[0.4cm]
QCD CROSS SECTIONS
}}
\vfill  
{\bf      Roberto BONCIANI$^{(a)}$ \footnote{This work was supported by the 
European Union under contract HPRN-CT-2000-00149.},}
{\bf      Stefano CATANI$^{(b)}$,
          Michelangelo L. MANGANO$^{(c)}$                               
}
     and 
{\bf     Paolo NASON$^{(d)}$
}\\[1cm]

{$^{(a)}$ Physikalisches Institut, Albert-Ludwig-Universit\"at Freiburg,
D-79104 Freiburg, Germany} \\
{$^{(b)}$ INFN, Sezione di Firenze, I-50019 Sesto Fiorentino, Florence, Italy
} \\
{$^{(c)}$ CERN, Theoretical Physics Division, CH~1211 Geneva 23, Switzerland} 
\\
{$^{(d)}$ INFN, Sezione di Milano, I-20136 Milan, Italy} \\
\end{center}                             
\nopagebreak
\vfill
\begin{abstract} 

We present the general expressions for the resummation,
up to next-to-leading logarithmic accuracy, of Sudakov-type logarithms
in processes with 
an arbirtrary number of hard-scattering partons.
These results document the
formulae used by the authors in several previous phenomenological
studies. The resummation formulae presented here, which are valid for 
phase-space
factorizable observables, determine the resummation correction in a
process-independent fashion. All process dependence is encoded in the
colour and flavour structure of the leading order and virtual one-loop
amplitudes, and in Sudakov weights associated to 
the cross section kinematics.
We explicitly illustrate the application to the case
of Drell--Yan and prompt-photon production.

\end{abstract}                                                
\vskip 1cm
CERN-TH/2003-125\hfill \\
June 2003     
\vfill       
\end{titlepage}

The perturbative QCD calculations of a large class of infrared and collinear 
safe observables are sensitive to Sudakov effects. Some classical examples of 
these observables are the $\ee$ energy--energy correlation in the
back-to-back region \cite{eec}, the cross section for Drell--Yan production
of lepton pairs in hadron collisions \cite{dy}, and several $\ee$ hadronic
event shapes in nearly two-jet configurations \cite{Catani:1992ua}. 

The Sudakov effects appear
when the observable is defined and/or measured close to the exclusive boundary
of its phase space. We generically denote by $y$ $(y > 0)$ the kinematical
variable that measures the distance from the exclusive boundary, so that the
Sudakov region is specified by $y \ll 1$.
When Sudakov sensitive
observable are computed as power series expansions in the QCD coupling $\as$,
the perturbative series involves terms of the type $\as^n L^{k}$ 
$(k\leq 2n)$, where $L=- \ln y$. These double logarithmic terms
are due to final-state radiation of soft and collinear partons, and are
a distinctive feature
of any short-distance dynamics that is 
governed by an underlying gauge field theory.
Since $L \gg 1$,
the presence of logarithmically-enhanced terms spoils the convergence of the
fixed-order expansion in $\as(Q^2)$, even if the observable is controlled
by a typical hard-scattering scale $Q$ whose value is large 
(such that $\as(Q^2) \ll 1$). The predictivity of perturbative QCD can be
recovered by reorganizing the perturbative series according to the degree
of divergence of the various logarithmic terms, and then 
by performing a systematic resummation, to all orders in $\as$, of the
contributions that are leading-logarithmic (LL), next-to-leading logarithmic 
(NLL), and so forth.

Resummed calculations up to NLL accuracy are available for several production
cross sections in hadron collisions (see the list of references
in Sect.~5 of Ref.~\cite{Catani:2000jh}), and for many hadronic event shapes
in $\ee$ annihilation (see e.g. Refs.~\cite{Catani:1992ua,eees}) 
and in deep-inelastic lepton--hadron scattering (see e.g. Refs.~\cite{dises}).
The inclusion of resummed Sudakov effects increases the theoretical accuracy 
of perturbative calculations, by extending their applicability to wider
phase-space regions and reducing the uncertainty coming from yet uncalculated
higher-order terms. This brings about relevant improvements
in phenomenological applications, as shown by the studies carried out in
recent years \cite{Frixione:2002kn}.
For example, in $e^+e^-$ annihilation the use of resummed calculations
has become the standard procedure in the comparison with data 
on hadronic event shapes \cite{Bethke:2000ai}: these calculations
allow one to extend the perturbative treatment towards the
two-jet region where statistics is higher; they also allow investigations
of hadronic physics at the interface between perturbative and non-perturbative
phenomena. In hadron collisions, resummed calculations often lead to
a considerable reduction in the scale dependence of the perturbative
predictions, as in the case of top quark production at the
Tevatron and bottom quark production at HERA~B 
\cite{bonciani,bcmn,Cacciari:2003fi}.

In recent years, different groups (KLOS \cite{klos}, 
BCMN \cite{bonciani,bcmn}, BSZ \cite{bsz}) have been working to develop
resummation formalisms that are process-independent and observable-independent.
The aim is to obtain generalized
resummation formulae that depend on universal coefficients, and 
that are applicable to different hard-scattering processes and different
classes of observables within the same process in terms of a minimal amount 
of information on the specific observable to be computed.
We have explicitly checked that our generalized 
resummation formulae (which are presented here) reproduce known NLL results 
for several quantities,
such as the thrust \cite{Catani:1991kz,Catani:1992ua} and $C$-parameter 
\cite{Catani:1998sf} distributions in $\ee$ annihilation,
the cross sections for the production of lepton pairs, vector bosons
\cite{dy} and Higgs bosons \cite{Catani:2001ic} in hadron collisions,
the structure functions \cite{Catani:1990rr,Contopanagos:1996nh}
in deep-inelastic lepton--hadron scattering at large values of the 
Bjorken variable. 
We used this formalism 
to derive the NLL resummed results of 
Ref.~\cite{bcmn} for the production of heavy quarks and prompt photons 
in hadron collisions. 
However, a general description of 
the 
formalism has never appeared
in the literature. The purpose of this work is to fill this gap.
Here we only give a brief illustration of our generalized resummation 
formulae.
More details on the formalism and its derivation are given in a forthcoming
paper.

The paper is organized as follows. We first consider QCD hard-scattering
processes without hadrons in the initial state. We discuss the kinematic
properties of the observables to which our resummation formalism applies.
Then, we present our generalized resummation formula up to NLL accuracy.
The explicit formula is expressed in terms of factorized final-state factors 
$(J_i)$ and interference terms $({\bf \Delta}^{(\rm int.)})$.
Then, we discuss the more general case of hard scattering in hadron
collisions and in processes with tagged hadrons in the final state.
Here the corresponding resummation formulae include additional
initial- and final-state factors $(\Delta_i)$.
We briefly illustrate the application of the general formalism 
by sketching the derivation of the resummation formulae presented in
Ref.~\cite{bcmn}.
Finally, we summarize our main results.

We begin our presentation by considering a generic infrared- and 
collinear-safe cross section $\sigma$ (or a related observable) in a
hard process that does not involve hadrons in the initial state
(for instance, hadron production in lepton collisions or in heavy-boson decays).
We suppose that the calculation of $\sigma$ at the leading order (LO) in 
QCD perturbation theory involves $m$ final-state QCD partons with four-momenta
$\{p_i\}=p_1, \dots, p_m$. For simplicity of presentation, we also limit
ourselves to considering the case of massless ($p_i^2=0$)
QCD partons (quarks, antiquarks and gluons).
Using a shorthand notation, we write the LO contribution $\sigma^{(LO)}$
to the cross section as
\beq
\label{siglo}
\sigma^{(LO)} = \int \;\rd\Phi(\sigma;\{p_i\}) \;\; 
| M^{(LO)}(\{p_i\}) |^2 \;,
\eeq
where $M^{(LO)}$ is the corresponding LO matrix element, and 
$| M^{(LO)} |^2$ denotes the squared matrix element summed over the colours 
and spins of the final-state QCD partons. The kinematics of the cross section
are fully described by the phase-space factor $\rd\Phi(\sigma;\{p_i\})$.
It includes the phase-space contributions for the production of the final-state
particles as well as any additional kinematics information (definition of jets,
event shapes, energy flows, ...) that is necessary to precisely define the
cross section $\sigma$ that we want to evaluate. The phase-space
dependence on $\sigma$ is briefly indicated by the notation $\rd\Phi(\sigma)$.
In particular, $\rd\Phi(\sigma)$ depends on the generic kinematic variable
$y$ that controls the distance from the Sudakov region. We assume that the LO
term $\sigma^{(LO)}$ is well-behaved\footnote{In practice, we consider the case
in which all the LO invariants $p_ip_j$ are of the order of the hard scale
$Q^2$ when $y \to 0$.}
(not singular) as $y \to 0$, 
while higher-order terms contain logarithmically-enhanced contributions of
relative order $\as^n L^{2n}$. The dependence of $\sigma$ on the momenta
of non QCD partons ($\gamma, Z^0, W^\pm, H, \dots$) is always understood.

Note that the Sudakov logarithms in $\sigma$ do not necessarily occur
by approaching the true physical phase-space boundary. These logarithms
can also appear inside the phase space of certain observables\footnote{A 
notable example in $e^+e^-$
annihilation is the $C$-parameter distribution, which has Sudakov logarithms in
the vicinity of $C=3/4$ \cite{Catani:1997xc}.
Other examples are discussed, for instance, in Refs.~\cite{dises,shoulder}.}.
Indeed, logarithmically-enhanced terms may arise \cite{Catani:1997xc} also 
if the phase-space boundary for a certain number of partons lies inside that
for a larger number, or if the observable itself is defined in a non-smooth 
way at some perturbative orders. In these cases, 
$\sigma^{(LO)}$ in Eq.~(\ref{siglo}) has to be
regarded as the lowest-order contribution at which those partonic boundaries
appear.

The practical feasibility of performing the resummation of the Sudakov 
logarithms at all perturbative orders depends on the capability of properly
approximating the higher-order contributions to $\sigma$. The approximation
regards both the QCD dynamics (i.e. the matrix elements)
and the cross section kinematics.
As for dynamics, since the Sudakov limit singles out multiple
radiation of soft and collinear partons, we can exploit the universal
(process-independent) factorization properties of the QCD multiparton
matrix elements in the infrared (soft and collinear) region (see e.g. 
Refs.~\cite{BCM} and \cite{Catani:dp}).
As for kinematics, we restrict our study to a (large) class
of observables, whose phase space is {\em factorizable}. By phase-space
factorization we precisely mean the following. At higher perturbative orders,
we consider the contribution to $\sigma$ from
the final-state radiation of additional partons with momenta
$q_1, \dots, q_k$. In the Sudakov limit $y \to 0$, these momenta are 
kinematically forced to
become soft or collinear to the momenta $\{p_i\}$, and the cross section
$\sigma$ is called factorizable if the corresponding phase space
$\rd\Phi(\sigma;p_1, \dots, p_m, q_1, \dots, q_k)$ behaves as
\beq
\label{psfact}
\rd\Phi(\sigma;p_1, \dots, p_m, q_1, \dots, q_k)
\ylim
\rd\Phi(\sigma;\{p_i\}) \; \left[ \rd q \right] \;\;
\prod_{j=1}^{k} \; u(\sigma, \{p_i\};q_j) \;\;,
\eeq
where $\rd\Phi(\sigma;\{p_i\})$ is the LO phase space, 
$\left[ \rd q \right]=\prod_{j} \rd^4 q_j \delta_+(q_j^2) /(2\pi)^3$ 
is the phase-space contribution from the unconstrained emission of 
the additional partons\footnote{We are treating the partons as distinguishable
particles. It the partons $j=1, \dots, k$ were identical, 
$\left[ \rd q \right]$ should be multiplied by a Bose-symmetry
factor of $1/k!$.} with on-shell momenta $q_1, \dots, q_k$,
and on the right-hand side 
we have neglected relative corrections that vanish in the soft and 
collinear limit.
The function $u(\sigma, \{p_i\};q_j)$ is called Sudakov weight. It depends
on the kinematical definition of the cross section $\sigma$ (such dependence
implicitly embodies the dependence on $y$), on the LO parton momenta
$\{p_i\}$ and on a {\em single} (soft and collinear) final-state momentum
$q_j$. The right-hand side of Eq.~(\ref{psfact}) implies that the kinematics
dependence on the soft and collinear momenta is {\em fully} factorized:
it is factorized with respect to the LO phase space and, moreover, there are 
no correlations between those momenta, since each Sudakov-weight factor 
depends on a single momentum $q_j$.

Note that the momenta $\{p_i\}$ on the right-hand side of 
Eq.~(\ref{psfact}) are not precisely the momenta of the LO partons on the
left-hand side. The former exactly coincide with the latter in the soft limit
$q_j \to 0$. When some of the momenta $q_j$ are not soft but collinear to the
momentum of one of the LO partons, say the parton~$i$, the momentum $p_i$
on the right-hand side is obtained by reabsorbing the longitudinal-momentum 
recoil produced by the collinear radiation.

As a consequence of the infrared and collineary safety
of $\sigma$, the Sudakov weight fulfils the following important property:
\beq
\label{ircolsaf}
u(\sigma, \{p_i\};q) = 1 \quad \quad {\rm when} \quad q=0 \;, \quad 
{\rm or} \quad q=(1-z)p_i  
\;\;\; {\rm for} \;\;\; i=1,\dots,m \;\;.
\eeq
Note that infrared and collinear safe observables are not necessarily
factorizable. A classical example of non-factorizable observables are jet 
rates when the jets are defined by the JADE jet-finder algorithm 
\cite{Brown:1990nm}.
Moreover, phase-space factorization is typically not achievable in the space of
the kinematic variables where the cross section is originally
defined. To overcome non-factorizable phase-space constraints, it is often
necessary to introduce a conjugate space. For instance, the constraints
of energy or transverse-momentum conservation are usually factorized by
respectively performing Mellin (or Laplace) or
Fourier transformations, and by working in the $N$-moment 
\cite{dy,Catani:1992ua,Catani:1996yz} or impact-parameter \cite{eec} space. 
Thus Eq.~(\ref{psfact}) can be valid either in the original space or 
in a properly defined conjugate space.
In the following, $y$ generically stands for either the original
Sudakov variable or the variable conjugate to it (more precisely, the inverse
of this conjugate variable) in the conjugate space.

To proceed further, we require one additional kinematics property on the
observable to be resummed. In the Sudakov limit, the energy flow accompanying 
the LO hard scattering has to be suppressed {\em uniformly} with respect to 
its emission direction. To be precise, we consider the behaviour  
of the Sudakov weight $u(q)$ (from now on, $u(\sigma, \{p_i\};q)$ is briefly
denoted by $u(q)$) when $y \to 0$ at fixed value of $q$. 
We write the on-shell 
four-momentum $q^\mu=\omega(1, {\bf {\hat q}})$ in terms
of its energy $q_0=\omega$ and a three-dimensional vector 
${\bf {\hat q}}$ of unit lenght $({\bf {\hat q}}^2=1)$, whose components
parametrize the emission angle. 
When $y \to 0$,  we thus require that
$u(q) \to 0$ and, without loss of generality, we can always assume, 
to dominant logarithmic accuracy, that $u(q)$ is approximable by a step
function:  
\beq
\label{dlogacc}
u(q) \simeq 
\Theta \left( \omega_{\rm max} - \omega \right) \;\;, \quad \quad
\omega_{\rm max}(y;\{p_i\},{\bf {\hat q}}) \ylim 0 \;\;,
\eeq
where, the upper bound $\omega_{\rm max}$ on the radiated energy depends on the
momenta $\{p_i\}$, on the emission angle ${\bf {\hat q}}$ and on the Sudakov
variable $y$. In the Sudakov limit, we then require
\beq
\label{uniform}
\frac{\rd \ln \omega_{\rm max}(y;\{p_i\},{\bf {\hat q}})}{\rd \ln y}
\yeq \lambda \;\;,
\eeq
where the power $\lambda$ is {\em positive}, independent of 
$y$ and $\{p_i\}$ and, in particular, {\em independent}\footnote{To be precise,
we allow Eqs.~(\ref{dlogacc}) and (\ref{uniform}) to be violated 
in angular regions of vanishing solid angle (for instance, when
$q$ is exactly parallel to a LO momentum $p_i$).}
of the radiation angle 
${\bf {\hat q}}$. Equations (\ref{dlogacc}) and (\ref{uniform}) state
in a formal way that, in the Sudakov limit,
the parametric suppression rate of the energy flow
emitted from the LO partons has to be uniform with respect to the radiation
angle.

Note that, by requiring the property 
in Eqs.~(\ref{dlogacc}) and (\ref{uniform}),
we exclude from our resummation treatment observables such as 
away-from-jet energy flows, and, in general, the so-called non-global
observables \cite{nonglob}.

The general kinematic properties of phase-space factorization
(see Eq.~(\ref{psfact})) and uniform suppression of the energy flow
(see Eqs.~(\ref{dlogacc}) and (\ref{uniform}))
are sufficient
to obtain our generalized resummation formula. The all-order cross section is 
generically denoted by $\sigma_u$. Here $\sigma_u$ can be either the original 
cross section $\sigma$, or the corresponding cross section in the 
conjugate space where
Eq.~(\ref{psfact}) applies (in this case, $\sigma$ is eventually computed 
by perfoming the inverse transformation of $\sigma_u$ to the original space).
The cross section $\sigma_u$ is written as
\beq
\label{sigu}
\sigma_u = \sigma_u^{(\rm res)} + \sigma_u^{(\rm fin)} \;\;,
\eeq
where the resummed component $\sigma_u^{(\rm res)}$ contains all the 
Sudakov logarithms, while $\sigma_u^{(\rm fin)}$ is well behaved (finite or
vanishing) order by order in $\as$ when $y \to 0$. Thus, 
$\sigma_u^{(\rm fin)}$ can reliably be evaluated by truncating its 
perturbative expansion at the first few perturbative orders. In practice,
$\sigma_u^{(\rm fin)}$ can be obtained from the fixed-order computation of
$\sigma_u$ by subtraction of the terms already included in $\sigma_u^{(\rm
res)}$ at the same fixed order.

The resummed component is given by
\beq
\label{sigres}
\sigma_u^{(\rm res)} = \int \;\rd\Phi_u(\sigma;\{p_i\}) \;\; 
| M^{(LO)}(\{p_i\}) |^2 \;\;\Sigma(u) \;,
\eeq
where $\rd\Phi_u(\sigma;\{p_i\})$ is either the LO phase
space $\rd\Phi(\sigma;\{p_i\})$, or its version in the conjugate space.
The expression (\ref{sigres}) is completely
analogous to the LO expression in Eq.~(\ref{siglo}). Sudakov resummation
is simply achieved starting from the LO result and performing the replacement
$| M^{(LO)}|^2 \to | M^{(LO)}|^2 \,\Sigma(u)$. The generalized effective form
factor $\Sigma(u)$ embodies the dependence on the Sudakov logarithms to all
perturbative orders. Since the Sudakov limit $y \to 0$
can formally be regarded as the limit $u \to 0$,
the presence
of logarithmically-enhanced terms in $\Sigma(u)$ is identified 
by contributions that order by order in $\as$ are formally divergent
when $u(q) \to 0$.
 
The Sudakov logarithms {\rm exponentiate}, and 
$\Sigma$ has the following 
structure
\beq
\label{sigmasim}
\Sigma \sim C(\as) 
\exp \{ {\cal G}(\as,L) \} = \left[ 1 + \as C_1 + \dots \right] \;
\exp \left\{ L \,g_1(\as L) + g_2(\as L) + \dots \right\} \;\;.
\eeq
The coefficient factor $C(\as)$ is independent of $u$ (or $y$)
and is due to hard virtual radiation.
Its perturbative coefficients $C_1, C_2, \dots$ depend on the specific
cross section $\sigma$, but, since they are not logarithmically enhanced,
they can be computed process by process at some finite perturbative orders.
The exponent ${\cal G}(\as,L)$ contains the logarithmically-enhanced terms.
The function $L \,g_1(\as L)$ resums the LL contributions $\as^n L^{n+1}$
in the exponent, the function $g_2(\as L)$ resums all the NLL contributions
$\as^n L^{n}$, and so forth\footnote{In our definition of LL, NLL, etc. terms,
we are referring to the logarithmic hierarchy of the various contributions 
to the exponent ${\cal G}(\as,L)$ (i.e. to $\ln \Sigma$)
in Eq.~(\ref{sigmasim}).
Our systematic resummation procedure thus differs from the ones that refer to 
the 
expansion of $\Sigma$ in successive logarithmic towers, such as
$\as^n L^{2n}$,  $\as^n L^{2n-1}$, etc.. In particular, our LL and NLL
contributions include more logarithmic terms
than those in the first two logarithmic towers of $\Sigma$ (see, for instance,
the discussion in Sect.~5 of Ref.~\cite{Catani:2000xk}).}.

The explicit NLL resummation formulae we are going to present
have the structure of Eq.~(\ref{sigmasim}), with the only difference that
the functions $C(\as)$ and ${\cal G}(\as,L)$ are matrices in the flavour and
colour indices of the hard-scattering partons, so that
exponentiation has to be understood in formal sense. The expression of the form
factor $\Sigma$ up to NLL accuracy is 
\beq
\label{sigformula}
\Sigma(u) = \left( \,\prod_{i=1}^{m} J_i(u) \right) 
\frac{\langle M_H(\{p_j\}) | {\bf \Delta}^{(\rm int.)}(u) 
| M_H(\{p_j\})\rangle} {| M^{(LO)}(\{p_j\}) |^2} \;\;.
\eeq

The first factor on the right-hand side of Eq.~(\ref{sigformula}) contains
all the LL terms and part of the NLL terms. It is given by the product of
the jet functions $J_i(u)$, and there is a jet function $J_i$
for each final-state parton $i$ in the corresponding LO process.
The function $J_i(u)$, which generalizes the jet function
of Refs.~\cite{dy,Catani:1992ua,bcmn},
embodies all the logarithmic terms produced
by multiple radiation of partons that are collinear (either soft or not)
to the direction of the momentum $p_i$ of the LO parton~$i$. 
The factorization in single-parton factors, $J_i$, 
is a consequence of the independent character 
(which follows from colour coherence) of QCD collinear radiation.

The remaining factor on the right-hand side of Eq.~(\ref{sigformula})
contains NLL terms and subleading logarithmic contributions. The
radiative factor ${\bf \Delta}^{(\rm int.)}(u)$ generalizes the 
analogous radiative factor introduced in the case of prompt-photon
hadroproduction \cite{bcmn}. It embodies all
the quantum-interference effects produced by non-collinear (large-angle)
soft-gluon radiation. In particular, this factor is sensitive to the
colour correlations due to the colour flow dynamics of the LO hard scattering.

The function $J_i(u)$ has the following resummed expression:
\beq
\label{jetfun}
\ln J_i(u) = 4\pi \int \frac{\rd^4 q}{(2\pi)^3} \,\delta_+(q^2)
\left( u(q) - 1 \right) \;
\frac{\Theta(z_{iq})}{p_iq} \;{\tilde A}\left(\as(2(1-z_{iq})p_iq)\right) 
\;P_i(z_{iq}) \;\;,
\eeq
where the functions $P_i(z)$, which depend on the flavour $(i=q,{\bar q},g)$
of the parton $i$,
are related to the Altarelli--Parisi splitting functions
\beq
P_{q}(z) = P_{\bar q}(z) = C_F \frac{1+z^2}{1-z} \;, \quad
P_{g}(z) = C_A \left[ \frac{2}{1-z} -2 + z(1-z) \right] + 
\frac{1}{2} N_f \left[ z^2 + (1-z)^2 \right] \;\;,
\eeq
and ${\tilde A}(\as)$ is the QCD coupling as defined in the bremsstrahlung 
scheme \cite{Catani:1990rr},
and is the related to the $\MSB$ coupling $\as$ by the NLO relation
\beq
{\tilde A}(\as) = \as \left[ 1 +\left( \frac{\as}{\pi} \right) \frac{K}{2} 
\right] \;, \quad 
K = C_A \left( \frac{67}{18} - \frac{\pi^2}{6} \right) 
- \frac{5}{9} N_f \;.
\eeq
The `energy fraction' $z_{iq}$ is defined with respect to a four-momentum 
$n_i^\mu$ as
\beq
z_{iq} = 1 - \frac{n_iq}{n_ip_i} \;.
\eeq
Indeed, to obtain our resummed formulae, we have introduced
$m$ auxiliary momenta $n_i^\mu$. These auxiliary momenta are arbitrary,
with the only constraint of being {\em time-like} $(n_i^2 > 0)$ and {\em
hard}\footnote{By hard, we mean that the invariants $n_i^2$ and $n_ip_j$
are of the order of the hard scale $Q^2$ when $y \to 0$.}. The LL terms in 
$J_i(u)$ do not depend on the definition of
these auxiliary momenta. The NLL dependence of $J_i(u)$ on $n_i$ is cancelled 
by the dependence on $n_i$ in the remaining factor on the right-hand side of
Eq.~(\ref{sigformula}), so that our expression for 
$\Sigma(u)$ is independent of $n_i$ up to corrections that are beyond the NLL
accuracy of the present formalism. The main motivation for introducing the
auxiliary momenta $n_i^\mu$ is to factorize collinear radiation in the
jet functions $J_i(u)$ without the introduction of explicit angular 
boundaries. In practical calculations, the definition of the momenta 
$n_i^\mu$ can be adjusted to simplify the evaluation of the integral 
in Eq.~(\ref{jetfun}).
 
The NLL contribution ${\bf \Delta}^{(\rm int.)}$ to the form factor is given by
\beq
\label{deltaint}
{\bf \Delta}^{(\rm int.)}(u) = {\bf {\overline V}}(u) \;{\bf V}(u) \;\;,
\eeq
where ${\bf V}(u)$ and ${\bf {\overline V}}(u)$ 
are matrices acting onto the colour indices of the LO partons.
The explicit expression of ${\bf V}(u)$ is
\beq
\label{vmatrix}
{\bf V}(u) = P_z \;\exp \left\{ \sum_{i\neq j} \int_0^1 \rd z 
\,\frac{
\langle u(q) \rangle
- 1}{1-z} 
\;\frac{\as((1-z)^2\mu_R^2)}{4\pi} \;\sum_c T_i^c \,T_j^c \, 
\ln \frac{4 (n_ip_i)^2 (n_jp_j)^2}{(p_ip_j)^2 \,n_i^2 \,n_j^2}
\right\} \;\;,
\eeq
where $\mu_R$ is the renormalization scale, to be chosen of the order
of the hard-scattering scale $Q$.
The sum $\sum_{i\neq j}$ runs over the labels $i,j=1,\dots, m$ of the LO
partons and $T_i^c, T_j^c$ ($c=1,\dots, N_c^2-1$) are the corresponding
colour charges: $T_i^c$ is the colour matrix\footnote{The colour charges are
defined according to the notation in Sect.~3.2 of Ref.~\cite{Catani:1996vz}.} 
in the fundamental (adjoint)
representation if the parton $i$ is a quark (gluon).
The operator $P_z$ denotes $z$-ordering in the formal expansion of the
exponential matrix, and  ${\bf {\overline V}}(u)$ is obtained from
Eq.~(\ref{vmatrix}) by simply replacing $P_z$ with ${\overline P}_z$,
the ordering operator that acts in the opposite order.
The notation $\langle u(q) \rangle$ in Eq.~(\ref{vmatrix}) stands for a 
properly defined average of the Sudakov weight $u(q)$. 
Parametrizing the light-like four-vector $q^\mu$ as
$q^\mu=\omega(1, {\bf {\hat q}})$,
the average is
performed over the angular directions ${\bf {\hat q}}$
at fixed value $q_0=\omega=(1-z)\mu_R$ of its energy $q_0$.
In practice, exploiting Eqs.~(\ref{dlogacc}) and (\ref{uniform}), and
neglecting terms beyond NLL accuracy, we simply have
\beq
\label{avunll}
\langle u(q) \rangle  = \Theta \left(y^\lambda - (1-z) \right) \;\;.
\eeq

In Eq.~(\ref{sigformula}), the NLL colour matrix ${\bf \Delta}^{(\rm int.)}$
acts onto the colour indices of the $m$-parton matrix element $M_H(\{p_i\})$.
This hard matrix element is independent of $u$ and is perturbatively
computable as a power series in $\as(\mu_R^2)$:
\beq
\label{hme}
M_H(\{p_i\}) = M^{(0)}(\{p_i\}) + \as(\mu_R^2) \; M_H^{(1)}(\{p_i\}) + 
{\cal O}(\as^2) \;\;,
\eeq
where $M^{(0)}$ is the LO matrix element and $M_H^{(1)}$ is the hard part of 
its one-loop virtual corrections. Since $M_H$ does not containt 
logarithmically-enhanced terms, it can be evaluated by truncation of its
perturbative expansion at some fixed order. In Eq.~(\ref{sigformula}), the 
dependence of $M_H$ on the colour indices is represented by the colour vectors
$|M_H\rangle$ and $\langle M_H |$, which are defined according to the notation
in Sect.~3.2 of Ref.~\cite{Catani:1996vz}.

The algebraic complications due to the non-trivial colour structure of the
NLL contributions are straightforwardly overcomed when the number $m$ of LO 
partons is
$m=2$ or $m=3$. In these cases the colour algebra can be carried out in
closed form, since the colour matrices $\sum_c T_i^c \,T_j^c$ in 
Eq.~(\ref{vmatrix}) are simply proportional to the unity matrix in colour 
space. The proportionality relations
are (see Appendix A in Ref.~\cite{Catani:1996vz})
\beeq
\label{colalg2}
\sum_c T_1^c \,T_2^c &=& - \,C_1 = - \,C_2  \quad \quad \quad \, (m=2) \;\;,\\
\label{colalg3}
\sum_c T_1^c \,T_2^c &=& \frac{1}{2} \left ( C_3 -  C_1 - C_2 \right)
\quad (m=3) \;\;,
\eeeq
where $C_i$ is the Casimir invariant of the parton $i$
($C_i=C_A$ if the parton $i$ is a gluon, $C_i=C_F$ if the parton $i$ is 
a quark or an antiquark). When $m=3$, $\sum_c T_2^c \,T_3^c$ and  
$\sum_c T_3^c \,T_1^c$ are obtained from Eq.~(\ref{colalg3}) by 
permutation of the parton indices $\{1,2,3\}$. In the case of processes with
$m \geq 4$ LO partons, the colour algebra cannot be handled without additional
information on the hard matrix element $M_H(\{p_i\})$,
since the colour can flow in many different ways through the
hard scattering. In general, the evaluation of $\sum_c T_i^c \,T_j^c$
(and of Eq.~(\ref{vmatrix})) requires the diagonalization of a linear 
combination
of $m(m-3)/2$ independent colour matrices 
(see Appendix A in Ref.~\cite{Catani:1996vz}).

The NLL resummed calculations of Ref.~\cite{klos}
deal with the non-trivial colour structure in the hard-scattering 
of $m=4$ LO partons by considering  `soft anomalous dimensions' of
gauge-dependent Wilson line operators. The Wilson line operators 
are introduced and properly defined on a process-dependent basis.
In this respect, the integrand in the exponent of Eq.~(\ref{vmatrix}) can be
regarded as a universal (process-independent and observable-independent) 
soft anomalous dimension matrix (in colour space), $\bf \Gamma$, of soft
non-collinear gluons radiated by  
hard scattering of an arbitrary number $(m \geq 4)$ of partons.
The explicit expression of $\bf \Gamma$,
\beq
\label{sadim}
\as \;{\bf \Gamma}(\{p_i, n_i\}) = \as \;\sum_{i\neq j}
\;\sum_c T_i^c \,T_j^c \, 
\ln \frac{4 (n_ip_i)^2 (n_jp_j)^2}{(p_ip_j)^2 \,n_i^2 \,n_j^2} \;\;,
\eeq
is gauge independent, though dependent on the auxiliary vectors $n_i$.
The Altarelli--Parisi splitting function $P_i(z)$ in Eq.~(\ref{jetfun})
controls the resummation of the Sudakov logarithms produced by collinear
(soft and hard) radiation, and, by analogy, $\bf \Gamma$ 
controls the resummation of the Sudakov logarithms produced by soft 
non-collinear radiation. Obviously, the definition of the boundary between 
the collinear and non-collinear regions is quite arbitrary. 
In Eq.~(\ref{sadim}), this arbitrariness is somehow parametrized by the
dependence on the auxiliary vectors $n_i$. For instance, using colour
conservation, $\sum_{i=1}^m T_i^c =0$ \cite{Catani:1996vz}, 
Eq.~(\ref{sadim}) can be rewritten as
\beq
\label{sadimmu}
\as \;{\bf \Gamma}(\{p_i, n_i\}) = \as \;\sum_{i\neq j}
\;\sum_c T_i^c \,T_j^c \, 
\ln \frac{4 \mu_i^2 \mu_j^2}{(p_ip_j)^2}
-  2 \,\as \;\sum_{i=1}^m C_i \;\ln \frac{(n_ip_i)^2}{n_i^2 \,\mu_i^2} \;\;\;,
\eeq
where $\mu_i$ are abitrary scales (e.g. $\mu_i=\mu_R$). The second term on the
right-hand side is proportional to the unity matrix in colour space, and
therefore it can be absorbed in a corresponding redefinition of the
jet functions $J_i$ in Eq.~(\ref{sigformula}).

We now consider the general case of cross sections in hadron collision
processes. In these processes, the hadronic cross section $\sigma$
is obtained by convoluting the partonic cross sections $\sigma_{a_1,a_2}$
with the parton densities $f_{a_1/h_1}(x_1,\mu_F^2)$ and 
$f_{a_2/h_2}(x_2,\mu_F^2)$
of the colliding hadrons with momenta $P_1$ and $P_2$:
\beq
\label{sighad}
\sigma = \sum_{a_1,a_2} \int_0^1 \;\rd x_1 \;\rd x_2 
\;f_{a_1/h_1}(x_1,\mu_F^2) \;f_{a_2/h_2}(x_2,\mu_F^2)
\;\sigma_{a_1,a_2} \;\;,  
\eeq
where $\mu_F$ is the factorization scale (to be chosen of the order
of the hard-scattering scale $Q$), and the sum $\sum_{a_1,a_2}$ runs over the 
flavours ($a_1,a_2=q,{\bar q},g$) of the incoming partons with momenta 
$p_1=x_1P_1$ and $p_2=x_2P_2$.

At the LO in QCD perturbation theory, the partonic cross section
$\sigma_{a_1,a_2}$ has the same structure as in Eq.~(\ref{siglo}), with
$m$ LO partons $i$ ($i=1,2$ are the incoming partons and
$i=3,\dots,m$ are the outgoing partons). The Sudakov logarithms at higher
orders can be produced by soft and collinear radiation emitted from either
the outgoing partons or the incoming partons in the LO hard scattering.
In general, to kinematically factorize the Sudakov effects in the parton 
densities from those in the partonic cross section, it is necessary to 
consider the $N$-moments (Mellin moments) $f_{a/h,N}(\mu_F^2)$ 
of the parton densities,
\beq
\label{npdf}
f_{a/h,N}(\mu_F^2) = \int_0^1 \;\rd x \;x^{N-1} \;f_{a/h}(x,\mu_F^2) \;\;,
\eeq
and the corresponding $N$-moments $\sigma_{a_1,a_2}^{N_1,N_2}$ of the partonic
cross section. 

In the following we limit ourselves\footnote{We thus exclude observables such
as, for example, the $Q_T$-distribution of Drell--Yan lepton pairs, where powers
of $\ln Q_T$ are produced also by radiation of hard quarks and gluons that are
collinear to the colliding partons.} 
to considering the cases in which 
{\em hard} radiation collinear to the incoming partons is suppressed in 
the Sudakov limit, so that the Sudakov logarithms are produced by soft
radiation (at any angles) and hard radiation collinear to the outgoing partons.
This simplifies the presentation of the resummed formulae, since the 
Sudakov effects do not change the flavours $a_1,a_2$ of the incoming partons.

Our NLL resummed formulae apply to the $N$-moments 
$\sigma_{a_1,a_2}^{N_1,N_2}$ of the partonic cross sections that fulfil the
factorization property of Eq.~(\ref{psfact}) in a properly defined conjugate 
space. 
We still require the properties in Eqs.~(\ref{dlogacc}) and 
(\ref{uniform}).
Infrared and collinear safety implies Eq.~(\ref{ircolsaf}) in the case
of radiation collinear to the outgoing partons $i=3,\dots,m$. When 
$i=1,2$ is an incoming parton, Eq.~(\ref{ircolsaf}) is modified
by a kinematical rescaling factor, which simply takes into account 
that we are considering the $N_i$-moments of the partonic cross section.
We have
\beq
\label{colfact}
u(q)= z^{N_i-1} \simeq \exp \{-(N_i-1) (1-z) \} 
\quad \quad {\rm when}  \quad q=(1-z)p_i 
\;\;\; {\rm for} \;\;\; i=1,2 \;\;,
\eeq
where the approximate equality is valid in the soft region $(1-z \ll 1)$
we are interested in.
The all-order partonic cross section has the same structure as in
Eq.~(\ref{sigu}), and its resummed component is obtained as in 
Eq.~(\ref{sigres}). The only difference is that the (final-state)
form factor $\Sigma(u)$
has to be replaced by a more general radiative factor 
$\Delta_{a_1,N_1;a_2,N_2}(u)$. The latter is given by an expression similar
to Eq.~(\ref{sigformula}), apart from two simple modifications that regard
the LL and NLL terms, respectively.

The modification of $\Sigma(u)$ at the LL level is obtained by performing
the 
replacement
\beq
\left( \,\prod_{i=1}^{m} J_i(u) \right) \longrightarrow 
\Delta_{a_1,N_1}(u) \;\Delta_{a_2,N_2}(u) 
\;\left( \,\prod_{i=3}^{m} J_i(u) \right) \;\;.
\eeq
In other words, $\Delta_{a_1,N_1;a_2,N_2}(u)$ is obtained from 
$\Sigma(u)$ by supplementing the product of the final-state jet functions 
$J_i(u)$ with an initial-state factor $\Delta_{a_i,N_i}(u)$ 
for each incoming parton $i=1,2$. The initial-state Sudakov factor 
$\Delta_{N_i}(u)$,
which generalizes the analogous $N$-moment factor of 
Refs.~\cite{dy,klos,bcmn},
resums the Sudakov logarithms produced by {\em soft} partons
emitted collinearly to the LO incoming parton $i=1,2$.
As in the case of the parton density $f_{a/h,N}(\mu_F^2)$, the radiative factor
$\Delta_{N_i}(u)$ depends on the factorization scale $\mu_F$ and on the
factorization scheme used to define the partonic cross section. The NLL
resummed expression of $\Delta_{N_i}(u)$ in the 
$\MSB$ factorization scheme is
\beeq
\ln \Delta_{a_i,N_i}(u) &=& 4\pi \int \frac{\rd^4 q}{(2\pi)^3} \,\delta_+(q^2)
\left( u(q) - u_i(z_{iq}) \right) \;
\frac{\Theta(z_{iq})}{p_iq} \;{\tilde A}\left(\as(2(1-z_{iq})p_iq)\right) 
\;P_i(z_{iq}) 
\nonumber
 \\
\label{pdfmatch}
&+& \frac{C_i}{\pi} \int_0^1 \,\rd z \; \frac{z^{N_i-1}-1}{1-z} 
\int_{\mu_F^2}^{4(1-z)^2(n_ip_i)^2/n_i^2} \frac{\rd k^2}{k^2}
\;{\tilde A}\left(\as(k^2)\right)\;\;.
\eeeq
The term on the right-hand side 
of the first line of Eq.~(\ref{pdfmatch})
is completely analogous
to the right-hand side of Eq.~(\ref{jetfun}) apart from the replacement
$( u(q)-1 )\to ( u(q)-u_i(z_{iq}))$, where $u_i(z)=u(q=(1-z)p_i)$.
The Sudakov limit typically
forces the parton distributions towards the large-$N$ (large-$x$) region,
and the term in Eq.~(\ref{pdfmatch})
matches the 
sensitivity of the parton distribution $f_{a_i/h,N_i}(\mu_F^2)$ to large
logarithms, $\ln N_i$, of the Mellin index $N_i$.

The modification of $\Sigma(u)$ at the NLL level regards the interference term 
${\bf \Delta}^{(\rm int.)}$ or, more precisely, its components
${\bf V}(u)$ and ${\bf {\overline V}}(u)$ in Eq.~(\ref{deltaint}).
Equation (\ref{vmatrix}) gives ${\bf V}(u)$ as an exponential of the
soft anomalous dimension matrix $\bf \Gamma$ in Eq.~(\ref{sadim}).
When going from processes with no initial-state partons to processes in
hadron--hadron collisions, we have to take into account that the parton momenta
$p_1$ and $p_2$ have to be crossed from the final to the initial state.
As for the anomalous dimension $\bf \Gamma$, the crossing simply amounts
to the following analytic continuation,
$\ln (p_ip_j) \to \ln (p_ip_j) + i \pi$, of the terms with
$i=1$ or $i=2$ and $j \geq 3$. Performing such a replacement, and using 
colour-charge conservation, $\sum_{j=3}^m T_i^c = - (T_1^c + T_2^c)$,
we obtain the overall replacement to be applied to $\bf \Gamma$:
\begin{equation}
\label{hhsadim}
{\bf \Gamma}(\{p_i, n_i\}) \longrightarrow
{\bf \Gamma}(\{p_i, n_i\}) \pm 4 i \pi \sum_c (T_1^c + T_2^c) (T_1^c + T_2^c)
\;\;.
\end{equation} 
Here, the signs $+$ and $-$ regard the replacements to be applied in the
evaluation of  ${\bf V}(u)$ and ${\bf {\overline V}}(u)$, respectively.
Note that the substitution on the right-hand side of Eq.~(\ref{hhsadim})
is effective only in the case of hadron--hadron collisions with $m \geq 4$
hard-scattering partons at LO. In fact, when $m=2$ we have 
$T_1^c + T_2^c=0$, and when $m=3$ we have 
$\sum_c (T_1^c + T_2^c) (T_1^c + T_2^c)= C_3^2$ so that the substitution
in Eq.~(\ref{hhsadim}) changes ${\bf V}(u)$ by a pure phase factor that is
cancelled by its complex conjugate phase factor in ${\bf {\overline V}}(u)$.

\setcounter{footnote}{0}

The replacement in Eq.~(\ref{hhsadim}), to be applied to the soft anomalous
dimensions in hadron--hadron collisions, has a direct and interesting
physical interpretation. As is well know from QED 
(see e.g. Ref.~\cite{Weinberg:mt}),
Coulomb-type virtual exchanges produce infrared-divergent Coulomb phases
that affect any scattering amplitudes. Being them phases, they cancel in the
evaluation of inclusive cross sections. The QCD analogue \cite{Catani:dp}
of the 
QED Coulomb phases are non-abelian Coulomb `phases', which are
colour matrices. They lead to non-abelian infrared divergences that cancel,
as in QED, 
in infrared- and collinear-safe observables with incoming massless partons
\cite{bncancellation,Catani:dp}.
However, in the non-abelian case, the cancellation mechanism of the 
infrared divergences can produce residual (and non-trivial) finite
contributions. The Sudakov logarithms produced by the $i\pi$-term
on the right-hand side of Eq.~(\ref{hhsadim}) are a manifestation of these
residual\footnote{The $i\pi$-term in Eq.~(\ref{hhsadim}) is not the absolute
overall effect of Coulomb-type interactions in hadron--hadron collisions.
It is the relative effect produced by the Coulomb-phase mismatch between processes with
no initial-states hadrons and processes in hadron--hadron collisions.
Thus, it has conveniently been introduced by the corresponding 
analytic continuation procedure.
} finite effects. Very soft (and, hence, infrared divergent) 
non-abelian Coulomb-type interactions do cancel in $\sigma$.
On the contrary, non-abelian Coulomb-type interactions of gluons
that are harder (and, hence, infrared finite) than 
bremsstrahlung gluons do not cancel, since they are trapped by the colour 
fluctuations produced by the radiated bremsstrahlung gluons.

We note that our resummation formulae apply also to processes in which
the partonic cross section has to be convoluted with the partonic
fragmentation functions of hadrons that are tagged in the final state.
The resummation formulae for these processes are simply obtained
by multiplicatively introducing a factor of $\Delta_{a_i,N_i}(u)$ for each
final-state parton $i$ whose momentum has to be convoluted with
the $N_i$ moment of the corresponding fragmentation function. The 
fragmentation factor $\Delta_{a_i,N_i}(u)$ is the same as for parton densities
(i.e. Eq.~(\ref{pdfmatch})),
provided the fragmentation functions are defined in the 
$\MSB$ factorization scheme.

To illustrate the use of our generalized resummation formulae,
we briefly sketch their application to the hadroproduction of Drell--Yan lepton
pairs \cite{dy} and of prompt photons \cite{bcmn}. 
In both cases, the hadronic
cross section is obtained by the factorization formula in Eq.~(\ref{sighad}),
and $\sqrt S$ ($S=(P_1+P_2)^2$) denotes the centre--of--mass energy. 
The hard-scattering scale is the invariant mass $Q$ of the lepton pair
in the first case, and the transverse energy $E_T$ of the photon in the second
case. 
The variable $y$ that parametrizes the distance from the Sudakov region is 
respectively given by $y= 1- Q^2/S$ and $y= 1- 4E_T^2/S$. We are interested in
the corresponding inclusive total (i.e. integrated over the rapidity of the
observed final state) cross sections in the Sudakov limit $y \ll 1$.
We thus consider the $N$-moments of the cross section in Eq.~(\ref{sighad}).
The $N$-moments are defined with respect to $(1-y)$ at fixed values of $Q$ and
$E_T$, respectively. This sets $N_1=N_2=N$ in the corresponding
partonic cross sections $\sigma_{a_1,a_2}^{N_1,N_2}$.
Since the moment variable $N$ is conjugate to $y$ through the Mellin
transformation, the Sudakov limit corresponds to $1/N \to 0$ (i.e. $N \to
\infty$). It is not difficult to prove that the factorization property
of Eq.~(\ref{psfact}) is valid in $N$-space for both processes.

In the Drell--Yan process, the LO hard-scattering subprocess is
$q(p_1) + {\bar q}(p_2) \to \ell \ell' (Q)$: the number of LO partons is 
$m=2$ and the LO kinematics set $Q^2=2p_1p_2$. 
The Sudakov weight is
\beq
\label{udy}
u_{DY}(q) = \exp \left\{ - N \frac{(p_1+p_2)q}{p_1p_2} \right\} \;\;,
\eeq
which fulfils Eqs.~(\ref{dlogacc}) and (\ref{uniform}) with $\lambda=1$.
Since the LO process has no final-state partons, the Sudakov resummation factor
$\Delta_{DY, N}$ is the product of ${\bf \Delta}^{(\rm int.)}$ and two
initial-state factors, $\Delta_{q_1,N}$ and $\Delta_{{\bar q}_2,N}$.
To apply our generalized resummation formulae, it is convenient to choose the
auxiliary vectors as $n_1=n_2=p_1+p_2$. We thus have 
${\bf \Delta}^{(\rm int.)}=1$ (see Eqs.~(\ref{deltaint}) and (\ref{vmatrix}))
and $\Delta_{q_1,N} = \Delta_{{\bar q}_2,N}$ 
(see Eq.~(\ref{pdfmatch})).
Moreover, since $u(q)=u_i(z_{iq})$, the term on the
right-hand side 
of the first line of Eq.~(\ref{pdfmatch})
vanishes. 
Therefore, from the second line of Eq.~(\ref{pdfmatch})
we finally obtain the known NLL result
for the Drell--Yan process \cite{dy}.

In the prompt-photon process
(see the second paper in Ref.~\cite{bcmn}), 
there are two independent LO hard-scattering subprocesses,
$a_1(p_1) + a_2(p_2) \to a_3(p_3) + \gamma(p_\gamma)$, with 
$\{ a_1=q, a_2={\bar q}, a_3=g \}$ and $\{ a_1=q, a_2=g, a_3=q \}$. 
The number of LO partons is 
$m=3$ and, in the Sudakov limit the LO kinematics set $2 E_T^2 \simeq 2p_1p_3
\simeq 2p_2p_3 \simeq p_1p_2$.
The Sudakov weight is
\beq
\label{upp}
u_{\gamma}(q) = \exp \left\{ - N \frac{2p_3q}{p_1p_2} \right\} \;\;,
\eeq
which fulfils Eqs.~(\ref{dlogacc}) and (\ref{uniform}) with $\lambda=1$.
The Sudakov resummation factor 
$\Delta_{a_1a_3\to a_3\gamma, N}$
is the product of two
initial-state factors ($\Delta_{a_1,N}$ and $\Delta_{a_2,N}$), one final-state
factor ($J_{a_3}$) and the NLL factor ${\bf \Delta}^{(\rm int.)}$.
To explicitly evaluate these factors, it is convenient to choose the following
auxiliary vectors: $n_1=p_1+p_3$, $n_2=p_2+p_3$, $n_3=p_1+p_2+p_3$.
Using this choice, it is straightforward 
to check that the term on the right-hand side 
of the first line of Eq.~(\ref{pdfmatch})
gives subleading (beyond the NLL accuracy) contributions. As for the colour
algebra, we can simply use Eq.~(\ref{colalg3}).
Performing the phase-space integrals in
Eqs.~(\ref{jetfun}), (\ref{vmatrix}) and using Eq.~(\ref{pdfmatch}),
we obtain the explicit resummed results anticipated in Sect.~4.2 of the
the second paper in Ref.~\cite{bcmn} and implemented in the phenomenological
study of Ref.~\cite{Catani:1999hs}.

We have discussed a generalized formalism to perform the resummation
of Sudakov logarithms in QCD hard-scattering processes. We have presented
explicit resummation formulae up to NLL accuracy.
The formulae are observable-independent and process-independent.
The dependence on the observable is
universally encoded in the one-particle Sudakov weight $u(\sigma, \{p_i\};q)$.
The dependence on the process is completely specified by 
flavour, colour charge and kinematics of the LO partons,
the LO matrix element and its hard virtual corrections at one-loop order.
This is the minimal amount of process-dependent information that is necessary
in any calculations at fixed perturbative order.
Within a specific process,
the formalism is applicable to a large class of QCD observables
that are specified by some kinematic properties, 
such as phase-space factorizability.
Phase-space factorization has already been exploited in the literature to
perfom resummation of several observables in processes controlled by LO hard
scattering of two and three QCD partons. The extension of the NLL
resummation techniques to multiparton processes 
requires a formalism able to deal with the radiation pattern of non-collinear
soft gluons emitted by 
the colour flow dynamics of the underlying hard-scattering. 
Available NLL resummed calculations of some specific cross sections
in four-parton processes 
treat the colour flow dynamics on a process-dependent basis. 
As for processes with higher number of LO partons,
no NLL resummed calculation has been presented so far.
Our NLL formalism applies to arbitrary processes with any number of 
hard-scattering partons and with arbitrary colour flow dynamics.
This opens prospects of phenomenological applications to 
multijet events at present and future high-energy colliders.

\end{document}